\documentclass[11pt]{article}
\usepackage{times}
\usepackage{graphicx}
\usepackage{fancybox}

\newenvironment{fminipage}%
    {\begin{Sbox}\begin{minipage}}%
    {\end{minipage}\end{Sbox}\fbox{\TheSbox}}

\title{Hubble, Hubble's law and the expanding universe}

\author{J. S. Bagla \\
Harish-Chandra Research Institute \\
Chhatnag Road, Jhusi, Allahabad 211019, India \\
E-mail: jasjeet@hri.res.in}

\begin{document}

\maketitle

\begin{abstract}
Hubble's name is associated closely with the idea of an expanding universe as
he discovered the relation between the recession velocity and distances of
galaxies. 
Hubble also did a lot of pioneering work on the distribution of galaxies in
the universe. 
In this article we take a look at Hubble's law and discuss how it relates with
models of the universe. 
We also give a historical perspective of the discoveries that led to the
Hubble's law.
\end{abstract}

\section{Hubble's Law}

Edwin P. Hubble is best known for his discovery of the relationship between
the distance and radial velocities of galaxies. 
All models of the universe are based on this relationship, now known as
Hubble's law. 
Hubble found that the rate at which galaxies are receding from us is
proportional to the distance, $V \propto r$, and used observations to
determine the proportionality constant.
This constant is now called the Hubble's constant in his honour. 
\begin{displaymath}
V = H_0 r 
\end{displaymath}
This form for the relationship has important implications (see Box-1).

We step back a little and fill in some background before continuing with our
discussion of the Hubble's law.

Early part of the $20$th century saw considerable discussion and activity
focused on understanding the structure of our own galaxy. 
Eventually it was understood that our galaxy is a fairly large system with
around a hundred billion stars.
Our galaxy, or the Galaxy, is nearly $80,000$~light years across.  
Astronomers use a different unit, a parsec ($1$~parsec $= 3.26$~light years)
and the Galaxy is around $25$~kilo parsecs across.  
The Galaxy is shaped like a disk with stars highlighting spiral arms in the
disk, there is also a spheroidal bulge near the centre of the
Galaxy\footnote{An {\sl appam} is a good description of the shape of disk and
  bulge, though not in proportion.}. 
The disk is surrounded by a faint halo of stars and globular clusters, each
globular cluster is a tight group of stars and these may contain $10^3 - 10^6$
stars each. 
The Sun is around $8$~kpc from the centre in the disk.

Many other galaxies have been known for a long time, however it was not very
clear whether these are a part of our own galaxy or are similar systems
located very far away. 
Hubble provided first reliable determination of distances to these galaxies
and convincingly proved that these are large systems of stars in their own
right. 

Now we revert to our discussion of the Hubble's law.  
In the velocity distance relation, velocities are measured in km/s, distances
in millions of parsecs (Mpc) and for this reason the Hubble's constant is
written in complicated looking units of km/s/Mpc even though it has dimensions
of inverse time.
We can recast the Hubble's law and write it in terms of direct observables. 
We do this step by step.
We first rewrite the recession velocity in terms of the redshift of spectral
lines that is determined directly from spectra. 
\begin{displaymath}
z = \frac{V}{c} = \frac{r}{c H_0^{-1}}  
\end{displaymath}
Speed of light is denoted by the usual symbol $c$.
Here $z$ is the Doppler redshift, note that this definition of redshift is
valid only for $|{\mathbf V}|/c \ll 1$.
Distances are often measured using reference stars, or other objects that are
known to have a given luminosity (see Box-2).
In such a case, the flux observed from the reference object is related to the
luminosity and the distance.
Energy emitted in a unit time is radiated uniformly in all directions and
eventually spreads out in a spherical shell of radius $r$.  
Energy observed per unit area, per unit time can then be written as
\begin{displaymath}
S = \frac{L}{4\pi r^2} 
\end{displaymath}
Here $S$ is the observed flux and $L$ is the luminosity.
If we observe a number of sources then the redshifts and observed fluxes are
expected to have the following relationship:
\begin{displaymath}
\log{z} = \log{\frac{r}{cH_0^{-1}}} = \log{\sqrt{\frac{L}{4 \pi S}}
      \frac{1}{cH_0^{-1}}} 
\end{displaymath}
In short, $\log{z} \propto -0.5 \log{S}$, where the constant of
proportionality depends on the Luminosity of the source and the value of the
Hubble's constant.
Astronomers use an inverse logarithmic scale called magnitudes to quantify
observed fluxes.
These are defined as:
\begin{displaymath}
m = - 2.5 \log{S/S_0}
\end{displaymath}
Here, $S_0$ is a reference flux.
Thus the relationship between redshift and magnitudes for a standard candle
is:
\begin{displaymath}
\log{z} \propto 0.2 m  ~~~~~~~ \Rightarrow ~~~ m \propto 5 \log{z}
\end{displaymath}
The full relation, with constants can be written as:
\begin{displaymath}
m - M =  5 \log{\frac{r}{10~{\rm pc}}} = 5 \log{\frac{r}{1~{\rm Mpc}}} + 25 = 5
\log{z} + 5 \log{\frac{cH_0^{-1}}{1~{\rm Mpc}}} + 25
\end{displaymath}
Here $M$ is the absolute magnitude that is defined as the magnitude if the
source is located at a distance of $10$~pc.  
Clearly, this is related to the luminosity and we note that unless we know the
luminosity of the source we cannot use the two observables (magnitude $m$ and
redshift $z$) to determine the Hubble's constant.

Hubble used a variety of ways to determine distances to
galaxies\cite{1926ApJ....64..321H}.   
Using the $100''$ Hooker telescope, he was able to find what are called Cepheid
variables in nearby galaxies. 
Cepheid variables are stars whose luminosity varies with time in a predictable
manner and the time period of variation is related to the average luminosity. 
This allowed Hubble to show that these galaxies were very far away, certainly
at distances that are much larger than the size of our own galaxy.
This was the first convincing determination of distances to other galaxies and
this is amongst Hubble's most significant scientific discoveries.
Box-3 illustrates the difficulties involved in measuring distances to galaxies
and measurement of Hubble's constant.

Hubble then looked for potential standard candles in order to be able to
determine distances to more distant galaxies.
He used brightest stars, brightest globular clusters, and many other sources
to determine distances to tens of galaxies.
Redshifts, and hence recession velocities of a number of galaxies were already
known from the work done by V. M. Slipher\cite{1917Obs....40..304S}.
The stage was now set for discovery of the distance-velocity relation.
Hubble approached this issue three years after his paper giving distances to
galaxies, and showed that the relationship was linear.
By this time, at least two relativists had already used the data
published by Hubble and Slipher to arrive at the same conclusion (see Box-4 for
details). 
However, it is easy to see that Hubble approached the problem from a different
perspective and that his discovery was made independently. 
The reason we associate this discovery with Hubble more than anyone is that he
continued to work on the problem and refine measurements in order to improve
determination of distances to other galaxies and to make a convincing case the
expansion was cosmological in
origin\cite{1931ApJ....74...43H,1935ApJ....82..302H}.  
On the other hand, Lema\^itre\cite{lem_hub,lem_hubt} and
Robertson\cite{rob_hub} were checking whether the velocity-distance
relationship expected from theoretical models was seen in nature or not.

The value for Hubble's constant determined by Hubble and others was around
$500$~km/s/Mpc. 
In the context of cosmological models, this indicated an age of the universe
around $2$~billion years\footnote{$1/H_0$ is a useful order of magnitude
  estimate of the age of the universe.  A more precise value requires
  knowledge of other parameters but it can be shown that the order of
  magnitude estimate is accurate to better than $50\%$.}. 
However, radio-active dating showed that some rocks on Earth were much older
than this. 
This led to a crisis as the universe must be older than all its contents. 
The problem was related to cross-calibration and incorrect identification of
some standard candles. 
It required painstaking work over the next quarter of a century to understand
these issues and get a better value for the Hubble's constant.
At present the value of Hubble's constant is
determined\cite{2000ApJ...529..786M} to be close to $70$~km/s/Mpc and the
corresponding age of the universe is $13.6$~billion years.  

The current challenge is to extend the redshift-magnitude relationship to
larger distances as we can find out more about the universe. 
Indeed, this aspect does not require determination of the Hubble's constant
and hence the errors due to cross-calibration are not relevant.
Figure~1 shows the Hubble diagram for Supernovae of type Ia, the brightest
standard candle known to us. 
We see clearly that at low redshifts ($z \ll 1$), the data satisfies the
Hubble's law ($m - M \propto 5\log{z}$) very well. 
At larger redshifts, the effects of space-time curvature modify this
relationship and hence we see some deviations\footnote{These deviations from
  Hubble's law at high redshift are used to constrain the contents of the
  universe.  The data shown here can be used to demonstrate the existence of
  an exotic component called dark energy that has negative pressure, and is
  leading to an accelerated expansion of the universe.}.
For these objects, we do not have a good calibration of the luminosity and
therefore we cannot use this data to determine Hubble's constant, but we can
verify the Hubble's law with the data.

\begin{figure}
\begin{center}
\includegraphics[width=10cm]{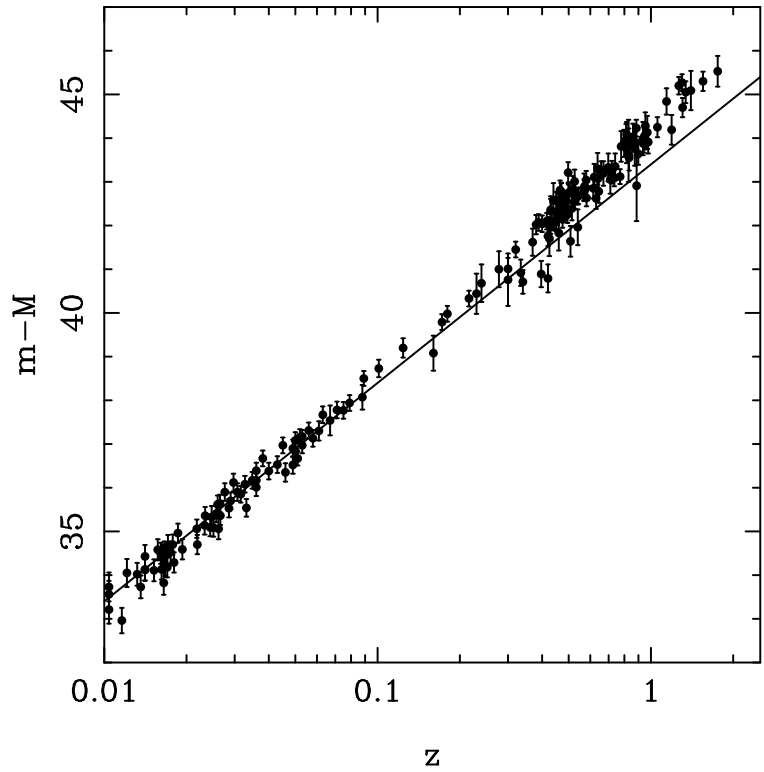}
\caption{Hubble diagram for Supernovae of type Ia.  The data shown here
  corresponds to the Gold+Silver sample\cite{2004ApJ...607..665R}.  The line
  corresponds to $m - M \propto 5 \log{z}$, the relation expected from
  Hubble's law.}
\end{center}
\end{figure} 

With this we end the story of Hubble and Hubble's law.  
What we have outlined here is only one aspect of Hubble's contribution. 
He contributed directly to many aspects of extra-galactic astronomy, indeed he
started this entire field.
Hubble convinced his employers about the need for larger telescopes and the
$100''$ Hooker telescope and the $200''$ Hale telescope at the Palomar
observatories that were set up primarily for extra-galactic work have been
used for research in all areas of astronomy.
I can only refer you to other articles in this volume for some details of
other contributions made by Edwin P. Hubble.

\begin{fminipage}[t]{12cm}
\centerline{\large\bf Box-1}

\bigskip

\centerline{\large\bf Hubble's Law and the Cosmological Principle}

\medskip

\noindent
The Hubble's law is written as 
\begin{displaymath}
V = H_0 r 
\end{displaymath}
with $V$ as the radial component of the velocity.  
The reason for writing down only the radial component is that this is the only
component of the motion that we can observe through the shift of spectral
lines. 
However, it is implicit in the form of the Hubble's law that the rate of
expansion is independent of direction.
In other words, the expansion is isotropic around us\footnote{An exception
  are anisotropic models\cite{1999toc..conf....1E} where galaxies recede from
  us at different rates in different directions.  However observations
  restrict the level of deviations from isotropy and one needs to construct
  models carefully in order to match observational data.}.
Clearly, the Hubble's law is {\it consistent}\/ with a vectorial relationship
between velocity and distance.
\begin{displaymath}
{\mathbf V} = H_0 {\mathbf r}
\end{displaymath}
In this form, it is easy to see that the Hubble's law retains its form if we
shift the origin: as seen from the origin, the galaxy at ${\mathbf r}_1$ recedes
with velocity ${\mathbf V}_1$.  
If we now try to rewrite the recession law in the frame of this galaxy, we
get:
\begin{displaymath}
{\mathbf r}' = {\mathbf r} - {\mathbf r}_1 
\end{displaymath}
\begin{displaymath}
{\mathbf V}' = {\mathbf V} - {\mathbf V}_1 = H_0 \left( {\mathbf r} - {\mathbf
    r}_1 \right) = H_0 {\mathbf r}'
\end{displaymath}
As claimed, the Hubble's law retains its form in the frame of any other galaxy
as well.  
Thus expansion appears the same in every direction, and from every place in
the universe. 
Unless we are observing at a special moment in the history of the universe,
this also means that the universe is homogeneous and isotropic.
The cosmological principle\cite{1917SPAW.......142E} elevates encapsulates
this idea, and it is noteworthy that a homogeneous and isotropic model of the
universe allows us to define a cosmic time.
Most models of the universe are based on this principle. 
\end{fminipage}

\begin{fminipage}[t]{12cm}
\centerline{\large\bf Box-2}

\bigskip

\centerline{\large\bf Distance Ladder}

\medskip

\noindent
Measuring distances to other galaxies is a challenging task as there is no
direct method of ascertaining the distance.  
There are two basic methods that are combined for measuring distances to
galaxies.
\begin{itemize}
\item
Parallax: We measure the parallax of nearby stars across Earth's orbit around
the Sun.  The parallax angle is $1'' (3.08 \times 10^{16}{\mathrm
  m}/r)$\footnote{The distance at which we get a parallax of $1"$ is called a
  parsec: $1$~parsec$= 3.08 \times 10^{16}$~m.}. 
Observations from the Earth can give reliable parallax measurements of up to
$0.1"$. 
\item
Standard Candle: If there is a source with known luminosity (energy output per
unit time), then the observed flux from such a source can be used to compute
the distance if the radiation from the source has not been absorbed by
intervening gas and dust.  Luminosity $L$, flux $S$ and distance $r$ are
related as $S = L / (4 \pi r^2)$ assuming that the source radiates uniformly
in all directions.
\end{itemize}
There are no standard candles where the luminosity is known {\sl a priori},
therefore one needs to do a calibration. 
In absence of such a calibration we can only measure relative distances and
not absolute distances.
Calibration is done by matching with the distance to a group of stars
measured using some other method, either parallax or some other standard
candle. 
A chain of standard candles is used and calibrated against each other, with
the nearest ones calibrated using the parallax method. 
This sequence of distance measurement methods is often referred to as the {\it
  distance ladder.} \/
Each step in the distance ladder involves cross-calibration and introduces
errors. 
The {\sl Hipparcos} space mission reduced errors by a significant amount by
providing accurate parallax measurements of up to $0.001"$, increasing the
number of stars with known parallax distances by a significant
factor\cite{1998AcHA....3..150S}. 
The upcoming Gaia mission\cite{2008AN....329..875J} is likely to make further
progress in this direction. 
\end{fminipage}

\begin{fminipage}[t]{12cm}
\centerline{\large\bf Box-3}

\bigskip

\centerline{\large\bf Why is measuring $H_0$ difficult?}

\medskip

\noindent
The main challenge in measuring the Hubble's constant is in accurate
determination of distances to galaxies\footnote{A noteworthy aside, it is
  apparent from the form of Hubble's law that it is possible to verify the
  relationship without any knowledge of the value of Hubble's constant.} 
This in turn requires us to use the distance ladder: the progression of
standard candles and parallax based methods of distance measurement through
cross-calibration. 
Each step in the distance ladder involves some uncertainty and hence adds to
the error in our knowledge of the luminosity of the standard candle that is
finally used.
If each step in the distance ladder introduces a few percent error, and the
standard candle used to determine distances to galaxies requires five steps of
cross-calibration then one can see that the error in luminosity of the
standard candle adds up to around ten percent. 

Another source of error are the observational uncertainties. 
If we are observing a star in a distant galaxy, then it is very difficult to
check if another star happens to be in the same direction. 
Finite resolution of imaging devices, the large surface density of stars in
galaxies, and, the large distances to galaxies makes {\sl blending} of stars
a very common phenomena.
The problem becomes more acute for distant galaxies as the projected density
of stars becomes large.
This introduces errors in the measured flux, and hence in the measured
distance. 

Lastly, galaxies are not merely receding from us due to expansion of the
universe, these also move around in the gravitational field of other
galaxies. 
This component of motion is called peculiar velocity.
The total velocities are thus:
\begin{displaymath}
{\mathbf V} = H_0 {\mathbf r} + {\mathbf v}
\end{displaymath}
Peculiar velocities are not expected have any average value, these are
expected to be random. 
If peculiar velocities have a typical magnitude $\sigma_v$, then these motions
introduce an error of order $\sigma_v / (\sqrt{3} H_0 r)$ in the determination
of the Hubble's constant where the factor of $\sqrt{3}$ arises from our use of
only one component of the peculiar velocity in determination of the Hubble's
constant. 
In order to appreciate the impact of this effect, let us consider some
numbers.
In our universe, $\sigma_p \sim 300$~km/s and $H_0 \simeq 70$~km/s/Mpc.  
Thus the factor $\sigma_v / (\sqrt{3} H_0 r) \sim 25\%$ at a distance of
$10$~Mpc, and the distance to the nearest large galaxy is less than $1$~Mpc. 
One can reduce this factor by measuring distances to a large number of
galaxies but that in itself is a fairly difficult and challenging task.
Therefore it is essential to use very distant galaxies for an accurate
determination of Hubble's constant.
On the other hand use of more distant galaxies increases errors due to the
first two factors mentioned above.
\end{fminipage}

\begin{fminipage}[t]{12cm}
\centerline{\large\bf Box-4}

\bigskip

\centerline{\large\bf Who discovered Hubble's law?}

\medskip

\noindent
The velocity distance relation encapsulated in Hubble's law\cite{hub_law} is
expected in all relativistic cosmological models, with the sole exception
being Einstein's static model.  
Dynamical models with expansion were discovered by Friedmann, Lema\^itre and
Robertson in the decade preceeding Hubble's discovery
At the time when Friedmann\cite{fr_met} published his solutions of Einstein's
equations, the size of the Galaxy was not known conclusively, nor was it known
whether other galaxies are a part of the Galaxy or are similar systems
situated at large distances.
The issue was finally settled by Hubble\cite{1926ApJ....64..321H} who used
Cepheid variables to measure distances to nearby galaxies and showed that
these are independent systems in their own right and lie at very large
distances. 
Recession velocities of a large number of galaxies had been measured
painstakingly over the years by V. M. Slipher\cite{1917Obs....40..304S}.
Thus the Hubble's law could have been discovered at any time after 1926. 
Lema\^itre\cite{lem_hub,lem_hubt} and Robertson\cite{rob_hub} discovered
cosmological solutions at this time, both realised that the recession of
galaxies constitutes an observational evidence of models of an expanding
universe.
Both used the data from Slipher and Hubble to verify the velocity distance
relation and determine the proportionality constant.
To them the form of the velocity-distance relation was natural and hence they
did not highlight it in their papers.
In an almost parallel effort, observers were trying to make sense of
relativistic models. 
Lundmark\cite{1925MNRAS..85..865L} decided to fit a quadratic relationship
between velocity and distance, postulating that there must be finite maximum
recession velocity. 
Hubble's paper\cite{hub_law} appears to be an attempt to demonstrate that the
quadratic term is either not required or that its coefficient must be very
small. 
By the time Hubble followed up this work with more
data\cite{1931ApJ....74...43H}, the language and the underlying paradigm
undergoes a significant shift: while the initial work by Hubble as well as
Lundmark's work uses methods common in stellar and galactic astronomy, there
is a sudden realisations of the cosmological scenario in all later work.

Hubble did indeed discover the Hubble's law, and did so independently, but he
was not the first one to get there.
\end{fminipage}

\end{document}